\documentclass[twocolumn,showpacs]{revtex4}
\usepackage{epsfig}

\begin{document}

\title{Density functional calculations for III-V diluted ferromagnetic semiconductors: A
Review} 

\author{Stefano~Sanvito}
\email{e-mail: ssanvito@mrl.ucsb.edu} 
\affiliation{Materials Department, University of California,
Santa Barbara, CA 93106, USA$\:$}

\author{Gerhard~Theurich}
\affiliation{Department of Chemistry, University of Pennsylvania, 
Philadelphia, PA 19104, USA} 

\author{Nicola~A.~Hill}
\affiliation{Materials Department, University of California,
Santa Barbara, CA 93106, USA} 

\date{\today}

\begin{abstract}
In this paper we review the latest achievements of density functional theory 
in understanding the physics of diluted magnetic semiconductors. 
We focus on transition metal doped III-V semiconductors, which show spontaneous 
ferromagnetic order at relatively high temperature and good structural compatibility 
with existing III-V devices.
We show that density functional theory is a very
powerful tool for i) studying the effects of local doping defects and disorder
on the magnetic properties of these materials, ii) predicting properties of new 
materials and iii) providing parameters, often not accessible from experiments, 
for use in model Hamiltonian calculations.
Such studies are facilitated by recent advances in numerical implementations 
of density functional theory, which make the study of systems with a very large 
number of atoms possible.
\end{abstract}


\maketitle

\section{Introduction}

The possibility of using the spin degree of freedom as well as the electronic charge
for electronic applications (``spintronics'') has received great
attention in the last few years \cite{Prinz}. A prototypical application of this
concept is the giant magnetoresistance effect (GMR) \cite{GMR}, where the
electrical resistance of a magnetic multilayer is changed by the application of
a magnetic field. GMR sensors are already commercially available for read heads 
in high density data storage devices, with performances increased by one order
of magnitude with respect to conventional sensors. However current GMR sensors do
not exploit completely the potential of spintronics. 
In existing electronic devices (for example personal computers) there are two 
main elements: the logic components and the data storage device.
The former are transistors based on semiconductor technology, 
while the latter is essentially a metallic magnetic film. 
Of course the ability to combine both logic elements and data storage in the 
same device will open completely new possibilities, with huge potential applications. 
Hybrid structures, where magnetic metals are used to inject spin electrons into
semiconductors have been shown to be problematic \cite{Rouk00,Lee99,Ham99}.
This is due to the large mismatch between the resistivities of semiconductors and 
metals, which seriously precludes effective spin-injection \cite{Schm1}. Although
this fundamental obstacle can be overcome, for instance by large contact
resistances \cite{mj}, it is natural to turn attention towards other directions 
and try to create all-semiconductor devices. These, of course, need magnetic
semiconductors.

At present several types of diluted magnetic semiconductors (DMSs) are available \cite{Ohno1}.
Among these, the DMSs based on III-V semiconductors are particularly important
because of their compatibility with existing III-V-based technology. III-V DMS
are obtained by low-temperature molecular beam epitaxy (MBE) deposition 
of III-V semiconductors with a transition metal such as Mn. The non-equilibrium 
growth is necessary to prevent the formation of additional phases and in general
only low concentrations of transition metal ions can be incorporated in the 
non-magnetic matrix. Nevertheless, despite the low
concentrations, the systems develop long range ferromagnetic order with
remarkably high Curie temperatures, $T_c$. For the known 
III-V-based DMS the highest Curie temperatures obtained are: $T_c\sim$~30$K$ for 
(In,Mn)As \cite{Ohno1,InMnAs}, $T_c$=110$K$ for (Ga,Mn)As \cite{Ohno1} and, 
very recently, a report of $T_c$=940$K$ for (Ga,Mn)N \cite{GaMnN}. Moreover, although this
paper is focused only on the III-V DMS, it is worth mentioning that ferromagnetism
with $T_c$ often above room temperature has been found in several other DMSs,
including Ge$_{1-x}$Mn$_x$ \cite{park1}, Cd$_{1-x}$Mn$_x$GeP$_2$ \cite{CdMnGeP}, 
Ti$_{1-x}$Co$_x$O$_2$ \cite{TiCoO} and Zn$_{1-x}$Co$_x$O \cite{ZnCoO}.

(Ga,Mn)As, although it has a $T_c$ far below room temperature, is the most widely
studied DMS. This is due to its structural compatibility with most epitaxially
grown III-Vs, which makes it ideal for building heterostructures and exploring
new device concepts. Examples that have been already achieved include
spin injection into heterostructures \cite{Aws1} and field-effect control of 
the ferromagnetism \cite{Ohno2}.

Three important features underlie the ferromagnetic order of (Ga,Mn)As,
and these are shared by the other Mn-doped III-V DMSs: 
i) Mn$^{2+}$ ions substitute the for Ga$^{3+}$ cations in the zincblende lattice 
providing localized spins ($S=5/2$ in Ga$_{1-x}$Mn$_x$As and In$_{1-x}$Mn$_x$As),
ii) there are free holes in the system although the actual concentration is much
smaller than the Mn concentration (despite the nominal valence suggests that the
two concentrations should be identical), iii) the hole spins couple
antiferromagnetically with the Mn spins, due to a dynamic $p$-$d$ coupling. 
Then the ferromagnetic behavior can be reasonably well described by the Zener 
model \cite{Dietl1}, in which antiferromagnetic exchange coupling
partially spin polarizes the holes, which in turn cause an
alignment of the local Mn spins. Within the Zener model, the interaction Hamiltonian 
between the hole spin $\vec{s}$ and the Mn spin $\vec{S}$ is
\begin{equation}
H=-N_0\beta \vec{s}\cdot\vec{S}\;,
\label{eq1}
\end{equation}
where $N_0$ is the concentration of the cation sites and $\beta$ is the $p$-$d$
exchange integral. $N_0\beta$ is usually called the exchange constant. 
If one simply uses the mean field approximation \cite{Dietl1,McD1}, in which 
the magnetizations of both carriers and Mn ions are considered to be uniform 
in space, we obtain an expression for $T_c$
\begin{equation}
T_c=
\frac{xN_0S(S+1)\beta^2\chi_s}{3k_B(g^*\mu_B)^2}\;,
\label{eq2}
\end{equation}
where $\chi_s$ is the magnetic susceptibility of the free carriers (holes in this
case), $g^*$ is their g-factor, $k_B$ is the Boltzmann constant and $\mu_B$ 
the Bohr magneton. 

The result of equation (\ref{eq2}) can be greatly refined by including a detailed
description of the band structure of the underlying non-magnetic semiconductors
or by incorporating correlation effects going beyond the mean field approximation. 
However, it is important to stress that the use of these models {\it always} 
requires parameters that are often difficult to extract from the experiments. 
For example the experimental value of the exchange constant $N_0\beta$
vary in the range 1-3~eV \cite{Szcz1,Ando1,Szcz2,Mats1,Omiy1,Oka1}. 

In addition, models based on any kind of mean field approximation naturally fail in
describing local effects, which occur on the atomic scale. The ferromagnetism 
of (Ga,Mn)As is very sensitive to the sample ``history'', such as the growth 
conditions \cite{Ohno1} and eventual after-growth processing \cite{Hay,Nitin}.
Since the growth dynamics certainly affects the microscopic configuration of the
samples, this suggests that knowledge of the local chemical environment is
crucial for understanding and modeling the properties correctly.

These considerations show that it is essential to have a microscopic theory 
providing information to the simpler mean field-like models.
Density functional theory (DFT) \cite{Kohn64,KS} is to date the most efficient
and accurate microscopic theory for describing the electronic, magnetic and 
structural properties of the ground state of electronic systems with a large 
number of degrees of freedom. 
Recent advances in the numerical implementations, mainly concentrated
in the use of improved pseudopotentials \cite{pseudo}, of efficient basis sets \cite{PAO} and
of order $N$ approximations \cite{Siesta1}, make possible the study of systems 
containing several hundreds of atoms. Such computational capabilities are 
required to study the DMSs in the low concentration limit.
The main aim of this paper is to provide a review of the achievements of
DFT in describing the properties of the DMSs. In particular we will show that
DFT is an invaluable tool for i) studying the effects of local doping defects 
and disorder on the magnetic properties, ii) predicting properties of new materials 
and iii) providing parameters, often not accessible from experiments, for use in
model Hamiltonian calculations.

The remainder of the paper is organized as follows. First we will briefly
overview the most recent developments in density functional theory.
In the following section we will
discuss the structural properties of (Ga,Mn)As and explain why zincblende MnAs
cannot be grown. Moreover we will also look at spin-orbit effects and explain
why this introduces only minor quantitative changes in the exchange constant
of (Ga,Mn)As. Then we will move to the low dilution limit, calculating the
exchange constants and discussing the limitations of mean-field models. In the
remaining sections we will consider the effects of the local microscopic
configuration of the Mn ions and possible intrinsic defects on the
ferromagnetism of (Ga,Mn)As. In particular we will look at the role of
intrinsic defects and at the transport properties of digital ferromagnetic
heterostructures (DFH) \cite{Kaw1}. Finally we will overview theoretical 
predictions for new materials and then we will conclude.

\section{Density Functional Theory}

Since its introduction in the 1960s \cite{Kohn64,KS} density functional
theory has evolved into a powerful tool that is widely used in
condensed matter theory and computational materials for the
calculation of electronic, magnetic and structural properties
of solids.
The method has been remarkably successful in predicting,
reproducing and/or explaining a wide variety of materials
phenomena. Specific examples range from early predictions
of phase transitions in silicon as a function of pressure \cite{Yin},
to determination of stable and meta-stable adsorption
geometries on metal surfaces \cite{Neug_Scheff} as well as many
successes in understanding the behavior of 
magnetic materials, including those described in this work.
 
The density functional formalism is based on the theorem that
for an interacting inhomogeneous electron gas in a static
external potential, $v({\bf r})$, there exists a universal
functional of the density, $F[\rho({\bf r})]$, independent of
$v({\bf r})$, such that the expression
\begin{equation}
E = \int v({\bf r}) \rho({\bf r}) d{\bf r} + F[\rho({\bf r})]
\end{equation}
has as its minimum value the correct ground state energy associated
with $v({\bf r})$ \cite{Kohn64}.
 
The true density, $\rho({\bf r})$, can in principle be exactly
obtained from the solution of an associated single-particle problem,
whose effective single-particle potential, $v_{eff}[\rho({\bf r})]$,
is a unique functional of $\rho({\bf r})$ \cite{KS}. As a
consequence, the many-electron ground state reduces to that of
a one-electron Schr{\" o}dinger equation:
\begin{equation}
\left[-\frac{1}{2}{\bf \nabla}^2 + v({\bf r}) + \int \frac{\rho({\bf r'})}
{\left| {\bf r} - {\bf r'}\right|} d{\bf r'} + \frac{\delta E_{xc}}
{\delta \rho({\bf r})} \right] \phi_i({\bf r}) = \varepsilon_i \phi_i({\bf r})
\end{equation}
where
\begin{equation}
\rho({\bf r}) = \sum \left| \phi_i \right|^2.
\end{equation}
The so-called Kohn-Sham wavefunctions, $\phi_i$, are single-particle
eigenfunctions that are strictly meaningful only
for determining $\rho({\bf r})$, and the Kohn-Sham eigenvalues,
$\varepsilon_i$, are the derivatives of the total energy with
respect to the occupation of state $i$. Note that they are
not strictly related to single particle excitation energies, although        
the Kohn-Sham band structure can sometimes be a useful tool
in the interpretation of photoemission (or similar) data.
 
For an arbitrary density there is no simple exact expression
for the exchange-correlation energy, $E_{xc}$, and so to
make progress, the so-called local density approximation
(LDA) is often made. Within the LDA, $E_{xc}$ is written as:
\begin{equation}
E_{xc}[\rho] = \int \rho({\bf r}) \epsilon _{xc}(\rho({\bf r})) d{\bf r}
\end{equation}
where $\epsilon_{xc}$ is the exchange-correlation energy per
electron of a uniform interacting electron gas of the same
density, $\rho$. The LDA is strictly valid only if $\rho({\bf r})$
is slowly varying, and many extensions exist which give improved
accuracy for systems with localized electrons. 
 
Density functional calculations for magnetic materials became
widespread in the late 1970s, with a number of studies of third
and fourth row transition metals \cite{Janak_Williams,Wang_Callaway,Moruzzi_77}.
These studies established that the local density
approximation gives results that are in reasonable agreement with
experiment for quantities such as cohesive energy, bulk modulus and
magnetic moments, provided that spin polarization is included explicitly,
by extending the LDA to the local spin density approximation (LSDA).
They also noted, however, that the calculated properties are very sensitive to
details of the structure and magnetic ordering, which can lead to
discrepancies between the LSDA results and experiment. The
most notorious of these is the well known prediction of the
incorrect ground state of iron (face centered cubic and antiferromagnetic,
rather than the correct body centered cubic and ferromagnetic)
by the LSDA.
 
A number of technical developments have facilitated the study of
magnetic materials, perhaps the most important being the
introduction of the fixed spin moment (FSM)
method \cite{Schwarz_Mohn,Moruzzi_FSM}. In the FSM method the ground
state of a constrained system with a fixed magnetic moment is
calculated. Not only this does speed convergence, but the total energy
{\it surface} in magnetic moment/volume space can be determined,
giving additional information particularly  about metastable magnetic
phases.  Also, implementation of Gaussian smearing \cite{Fu_Ho} and
related schemes have helped to speed convergence of calculations for magnetic
metals with partially filled $d$ bands and complex Fermi surfaces, in
which it is difficult to carry out integrals over the occupied part of
the Brillouin Zone.
 
In parallel with these technical developments, extensions and improvements
to the LSDA have also been explored. The usual generalized gradient (GGA)
and weighted density (WDA) approximations that give improved results for
non-magnetic systems do not tend to give systematic improvement for magnetic
materials, although the GGA does at least predict the correct ground state
for iron. For more information about these approximations see reference
\cite{David_Singhs_book} and references therein. Methods such as the
LDA+U \cite{anis1,anis2}, and self-interaction-correction \cite{pzu} are 
specifically tailored to treat strongly correlated systems, and therefore 
are more appropriate for magnetic systems with narrow $d$ or $f$ bands.       

\subsection{Codes available}

There are many excellent computer programs available today
for performing density functional theory calculations. These use a 
range of different methodologies, have different specialties, and 
are widely varying in cost (both in dollar amount and in their computer 
requirements). In this Section we describe some of the most popular
programs.

Many DFT programs are based on the so-called plane wave pseudopotential
(PWPP) method \cite{Yin}, in which the wavefunctions are expanded
in a plane-wave basis, and the electron-ion interaction is modeled
by a pseudopotential. Plane wave basis sets offer many advantages in DFT
calculations for solids, including completeness,
an unbiased representation, and arbitrarily good
convergence accuracy. Publically available PWPP codes include {\sc DoD} 
Planewave \cite{DOD} which is a general purpose scalable planewave basis 
density functional code that treats insulators, semiconductors, metals 
and magnetic materials,
{\sc ABINIT} \cite{ABINIT}, which allows both DFT and density functional
perturbation theory calculations, and
{\sc Spinor} \cite{Spinor} which extends the usual LSDA formalism to include
spin-orbit coupling and generalized non-collinear magnetism. Both {\sc ABINIT}
and {\sc Spinor} are PWPP codes and are available under the GNU General Public
License \cite{Gnu}.   
One of the most popular ``semi-commercial'' PWPP codes is the {\sc VASP}
package \cite{vasp} developed at the University of Vienna, Austria.
{\sc VASP} allows DFT and molecular dynamics calculations, and is quite fast
because of its use of ultra-soft pseudopotentials. The developers charge
a nominal fee for the source code, and require authorship on the first
publication using the code. 
There are also a number of fully commercialized density functional
codes that are targeted in large part at chemical and
pharmaceutical companies. For example accelrys \cite{Accelrys}
markets the PWPP {\sc CASTEP} code \cite{Castep}.  

For systems involving a large number of atoms in the unit cell, plane-wave based
DFT algorithms are not ideal because of the large computational overheads involved. 
For this purpose it is convenient to use codes based on localized atomic orbital 
basis sets, although their numerical implementation is usually quite complicated. 
Most of the results of this paper are obtained with the code {\sc siesta} 
\cite{Siesta2,Siesta3}, which combines pseudopotential techniques with a 
pseudoatomic orbital basis set \cite{PAO}. The code is highly optimized to deal 
with large systems, and both efficient order~N methods and molecular dynamics 
tools are available. The developers require an initial collaboration and 
co-authorship on works produced by {\sc siesta}. 

Traditionally however, magnetic materials have been studied using 
all-electron methods with mixed basis sets, such as the linear 
augmented plane wave (LAPW) \cite{David_Singhs_book}, linear muffin 
tin orbital (LMTO) \cite{LMTO} or Korringa-Kohn-Rostoker (KKR) \cite{KKR} 
approaches. 
Again many codes are available.
For example, the Vienna University of Technology produces
a full-potential LAPW code called {\sc WIEN97} \cite{Wien97} which is a highly
accurate, all-electron code that includes relativistic effects. A
small (\$350 at press time) fee is charged. 
The Stuttgart LMTO program \cite{Stuttgart_LMTO}
is a fast and efficient tool for calculation of charge- and
spin-self-consistent band structures, partial densities of states,
Fermi surfaces, total energies, and the partial pressures. In addition,
the program delivers tools for analyzing the electronic structure and
chemical bonding such as orbital-projected band structures, crystal
orbital Hamiltonian populations and electron localization functions.

\section{Zincblende M\lowercase{n}A\lowercase{s}}

As we mentioned in the introduction, so far Mn has been incorporated in GaAs 
only at concentrations smaller than 7\%. Considering that the Zener model 
predicts that $T_c$ increases with $x$ (see equation (\ref{eq2})), this poses a 
severe limit to the highest Curie temperature obtainable. 
It is therefore of great interest to investigate if
there are some conditions, which allow the growth of (Ga,Mn)As with higher Mn
concentration or even ultimately zincblende MnAs. In this Section we describe
DFT calculations from the literature that investigate both the stability and
the electronic and magnetic properties of the zincblende MnAs.

Several groups have calculated the electronic structure of zincblende MnAs
using both the LSDA \cite{shirai,ogawa,us1,shirai1} and the GGA \cite{silvia} 
approximations for the exchange correlation potential. The calculated 
lattice constant, $a_0$, for zincblende MnAs is found to be around 5.6-5.7~\AA. 
This result, also confirmed by relaxation calculation of (Ga,Mn)As \cite{jain}, 
is quite appealing since a hypothetical zincblende MnAs appears to have a lattice 
constant very close to that of GaAs ($a_0=5.6533$\AA). 
These results seem to conflict with the extrapolation to $x$=1 
(assuming Vegard's law) of the lattice constant of Ga$_{1-x}$Mn$_x$As and 
In$_{1-x}$Mn$_x$As \cite{Ohno1}, which give respectively $a_0$=5.98\AA\ and
$a_0$=6.01\AA. However it is worth noting that the lattice constant 
has been measured only for $x\ll1$ and that a linear extrapolation to $x=1$ may
be not valid. Moreover in real (Ga,Mn)As samples the presence of
intrinsic defects (mainly As antisites As$_\mathrm{Ga}$) can play an important 
role in determining the structural properties. This is clearly demonstrated in
low-temperature annealing experiments \cite{Hay,Nitin}, where the lattice
constants of samples with the same Mn concentration, change depending on
annealing conditions (temperature and time).

It is also important to point out that generally the LDA approximation
underestimates the equilibrium lattice constant, in particular if strong $p$-$d$
hybridization is present. This is probably the situation for zincblende MnAs,
though the good agreement with experiments for a different MnAs lattice structure
(the NiAs-type structure), gives confidence in the LSDA results. 
Furthermore a recent LDA+U calculation \cite{park} (LDA+U usually corrects the 
tendency to overbinding of the LDA) finds a lattice constant for MnAs very 
similar to that of GaAs \cite{minp}. This suggests that LDA provides a good
description of, at least, the structural properties.
Finally we must point out that a much larger lattice constant ($a_0$=5.9\AA) 
has been found by Shirai {\it et al.} \cite{shirai,shirai1} within the LSDA.
However this is probably due to an artifact of the minimization procedure 
used.

\begin{figure}[htbp]
\narrowtext
\epsfysize=6.7cm
\epsfxsize=8cm
\centerline{\epsffile{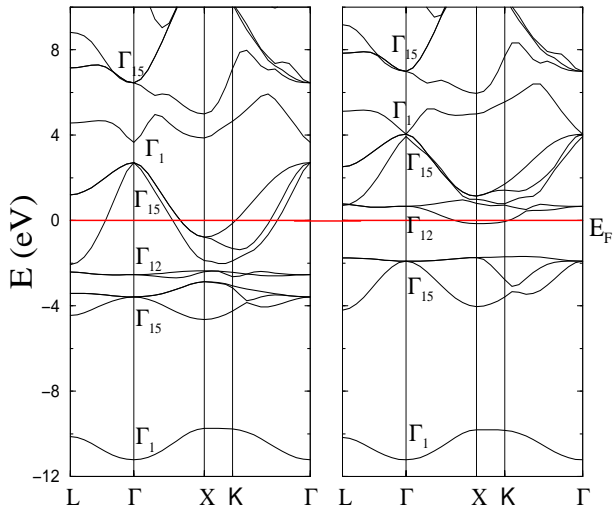}}
\caption{Band structure for zincblende MnAs at the LDA energy minimum ($a_0$=5.7
\AA ).
The figure on the left corresponds to the majority spin and the one on the 
right to the minority. The horizontal line denotes the position of the Fermi 
energy, which has been chosen to be 0~eV. }
\label{FIG1}
\end{figure}

Turning our attention to the electronic properties, in figure \ref{FIG1} we
present our calculated band structure for MnAs at the LDA equilibrium lattice constant
$a_0$=5.7\AA\ from reference \cite{us1}. The results are obtained with the code {\sc
Spinor} \cite{Spinor}.
Calculations by other groups at similar
lattice constants give similar band structures \cite{silvia}.
We first note that, excluding the presence of the Mn $d$ bands, the bandstructure
closely resembles that of the non-magnetic III-V semiconductors. If we consider
the majority band first, we can easily identify the As $p$ valence band (first
$\Gamma_{15}$ point above $E_{\mathrm F}$) and the first of the conduction 
bands (first $\Gamma_1$ point above $E_{\mathrm F}$). However the strong
interaction with the Mn $d$ states pushes the former toward higher
energies and they become half filled. The Mn $d$ bands, which are split into the
doubly degenerate $e$ band ($\Gamma_{12}$) and the triply degenerate $t_{2}$
band ($\Gamma_{15}$) are below the Fermi energy and entirely occupied. 
In contrast, in the minority band there is a large split between the 
$t_{2}$ and the $e$ states, which gives rise to a large gap in the band
structure. 

It is very important to note that at this lattice constant the Fermi energy in
the minority spin band cuts through the edge of the Mn-$d$ ($e$) states. This
suggests that a tiny expansion of the structure, reducing the Mn-$d$ ($e$) 
band width, will move the Fermi energy into the minority spin gap. In fact
we have predicted \cite{us1} that for a lattice constant larger than $a_0$=5.8\AA\
zincblende MnAs will be a half-metal. This result, confirmed by other calculations
\cite{shirai,ogawa,shirai1}, is very attractive since half-metallic systems are
sources of fully spin polarized currents. We therefore investigate if 
there are conditions under which such a zincblende phase of MnAs can be grown.

\subsection{MnAs: Zincblende vs NiAs-type structure}

\begin{figure}[htbp]
\narrowtext
\epsfysize=5.5cm
\epsfxsize=7cm
\centerline{\epsffile{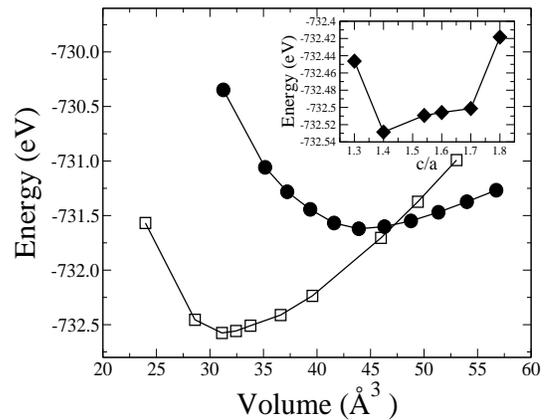}}
\caption{Total energy as a function of the MnAs pair
volume for the NiAs-type (squares) and the zincblende (circles) structure. 
Note the large stability of the NiAs-type structure over a very broad
volume range. In the inset we present the same quantity for the NiAs-type
structure as a function of $c/a$.}
\label{FIG2}
\end{figure}

The main obstacle to the growth of zincblende MnAs is its instability with 
respect to the NiAs structure. The NiAs-type structure is a
hexagonal structure (space group $P$6$_3$/$mmc$) with [6]-coordinated Mn. 
We have studied the relative stability of the NiAs-type and the zincblende
structures by comparing the total energy per MnAs pair as function of the unit
cell volume \cite{us1}. The results are presented in figure \ref{FIG2}.
It is clear that the NiAs-type structure has a much lower total energy and also a
denser lattice. Therefore it is the stable structure at all
thermodynamically accessible pressures. However we note that if the lattice is
forced to expand there is a crossover between the two structures, with the
zincblende being favorable for volumes larger than $\sim$47~\AA$^3$ per MnAs pair. 
This corresponds to a lattice constant for the zincblende structure of $a_0$=5.8~\AA, 
the same at which the transition to half-metal occurs. 
Therefore there is the hope that zincblende
MnAs could be grown if ``negative pressure'' were applied, for example if it were 
grown on a substrate with large lattice mismatch. Unfortunately this is also very 
challenging. 

In the insert of figure \ref{FIG2} we show the total energy as a function of 
the ratio between the two lattice constants of the NiAs-type structure 
($c/a$ ratio) at constant volume (the experimental volume). 
The figure shows that the NiAs-type structure can accommodate
large structural distortions without corresponding large energy costs (the total
energy changes of only about 20~meV when going from $c/a=1.7$ to $c/a=1.4$). 
This means that at equilibrium it is energetically more favorable for the 
system to distort the cell, instead of increasing the volume and inducing 
a NiAs-type to zincblende transition. 
In conclusion DFT calculations have shown that zincblende MnAs would indeed
have desirable properties but that it is unstable and will always tends to form the
less attractive NiAs phase.
Results analogous to those presented here are obtained for MnBi \cite{us1}, 
MnSb and MnP \cite{silvia}, suggesting that the low solubility limit of Mn 
is a characteristic of all the III-V's.

\subsection{The effect of Spin-orbit coupling}

The commonly used $p$-$d$ interaction Hamiltonian for dilute magnetic 
semiconductors is of the Kondo form given in equation (\ref{eq1}). 
The strength of the interaction is governed by the exchange constant 
$N_0\beta$. In this Section we will address the question of how $N_0\beta$ 
is affected by the spin--orbit coupling.

Despite the formal similarity of equation (\ref{eq1}) to a Heisenberg
exchange interaction the $p$-$d$ interaction in DMS materials does not
originate from the Coulomb interaction but arises as a result of the
hybridization between $p$ and $d$ derived bands in the crystal. The
appropriate Hamiltonian to describe transition metal impurities in
a host crystal is the well--known Anderson Hamiltonian.
However, Schrieffer and Wolff have shown that the Anderson Hamiltonian can
be transformed into a Kondo--like form \cite{schrieffer}, containing a 
term similar to that of equation (\ref{eq1}). This transformation
relates the effective exchange integral ($\beta$), which will be
negative in general, to the matrix elements of the interaction
potential between the bands of the crystal. We will discuss the
results of this section using the following model Hamiltonian for
the valence band states:
\begin{equation}
\label{eqGER1}
H = H_0 + H_X^{pd} + H_{soc}\ .
\end{equation}
Here $H_0$ is the crystal Hamiltonian without $p$-$d$ and without
spin--orbit interaction. $H_X^{pd}$ is the $p$-$d$ interaction given
by equation (\ref{eq1}) and $H_{soc}$ is the spin--orbit interaction.

Naturally, if the potential changes, the hybridization between the bands 
will be affected and thus the effective exchange integral will change
accordingly. Note that if $\beta$ were a ``real'' Coulomb exchange
integral it would not be directly affected by a change in the
potential. Only a very small indirect effect due to the change of
the self--consistent charge distribution would be expected in that
case.

In the (Ga,Mn)As crystal the As anions introduce a substantial
spin--orbit interaction which is about one order of magnitude
smaller than the exchange constants determined within
scalar--relativistic calculations. In state of the art
scalar--relativistic density functional calculations spin--orbit
coupling is taken into account only effectively by using averaged
potentials, which conserve all non--relativistic symmetry
properties of the electronic states in the crystal. However
in practice the symmetry
relations change drastically when spin--orbit coupling is
considered for a spin--polarized system.

There are two questions to be answered. First, is it still
possible to transform a hybridization interaction into the
Kondo--form when the potential becomes spin--dependent, as is the
case when spin--orbit coupling is included explicitly? Secondly, if
the Kondo--form still holds, how much will the exchange
constant change due to the change in the potential? In order to
answer these questions we performed density functional
calculations based on fully--relativistic pseudopotentials
\cite{theurich} for GaAs and MnAs, using the code {\sc Spinor}. 
Naturally the obtained density
functional results are fully self--consistent and do not lend
themselves easily to discuss individual terms of the model
Hamiltonian introduced in equation (\ref{eqGER1}) separately. We 
therefore proceed in the opposite direction and check whether the 
density functional results can be interpreted by the model Hamiltonian.

\begin{figure}[hbtp]
\centerline{\epsfig{file=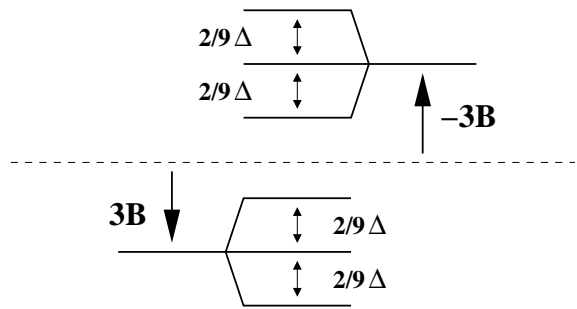,scale=.75,angle=0}} 
\caption{\small{Exchange and spin--orbit splitting of the valence 
band top in MnAs. The split of the center of the band is $6B$ where
$B=\frac{1}{6}N_0\beta\langle S_z\rangle$.}}
\label{FIGGER1}
\end{figure}

We assume that $H_0$ of equation (\ref{eqGER1}) has been solved for 
the valence band states at the Brillouin zone center. 
For scalar--relativistic band structures of zincblende semiconductors 
like GaAs the valence band top is six times degenerate.
This is also the case for the magnetic system MnAs as long as 
$H_X^{pd}$ is turned off. However,
when $H_X^{pd}$ is turned on the $p$-like states at the top of the
valence band will split according to their spin orientation into two groups of
triply degenerate states, separated by the energy
$N_0\beta\langle S_z\rangle$, following equation (1) (see also the discussion
of Section IV B). Here $\langle S_z \rangle$ is the
z-component of the spin--polarization per unit--cell. This is the
result found by the scalar--relativistic approach. Finally, treating
$H_{soc}$ as a perturbation, all remaining degeneracies are lifted
by first order energy corrections as is shown in figure \ref{FIGGER1}. Note
that the second order corrections are only of order 1~meV for
typical values of $\Delta$, the spin--orbit splitting of the
valence band at $\Gamma$, and $B=\frac{1}{6}N_0\beta\langle S_z\rangle$.

Now let's turn to the density functional results. Table \ref{TABGER1}
lists the the scalar-- and fully relativistic valence band edge
energies for MnAs.
The DFT calculations confirm the model discussed above in that all
scalar--relativistic degeneracies are lifted when spin--orbit
coupling is included explicitly. We also find that, in both groups,
one state lies approximately at the same energy as the
scalar--relativistic degeneracy, as predicted by the model. However, the energy
difference between the three minority states is approximately
50~meV, whereas the same number for the majority states is found to be
approximately 100~meV. According to figure \ref{FIGGER1} the model
predicts a splitting of $\frac{2}{9}\Delta$ for minority and
majority states alike, which is approximately 80~meV, using the
GaAs spin--orbit splitting for $\Delta$. Hence the model and
density functional results show significant discrepancies of about
30\% which would indicate that the model Hamiltonian of
equation (\ref{eqGER1}), and therefore the Kondo-form of the $p$-$d$
interaction, is questionable for MnAs when spin--orbit coupling is
taken into account explicitly. However, the deviations are about
two orders of magnitude smaller than the splitting between the
majority and minority states. Hence the corrections due to
spin--orbit coupling affect the exchange constant $N_0\beta$ for
MnAs only in the order of 1\%.

\begin{table}[hbtp]
\centerline{
\begin{tabular}{ccc}
\hline
& Minority (eV) & Majority (eV) \\ \hline \hline
without $H_{soc}$ & -2.405  & 2.405 \\ \hline \hline
with $H_{soc}$    & -2.355  & 2.499 \\ 
with $H_{soc}$    & -2.408  & 2.399 \\ 
with $H_{soc}$    & -2.458  & 2.308 \\ \hline \hline
\end{tabular}}
\caption{\small{Top of the valence band levels for MnAs at the Brillouin 
zone center determined by density functional method with and without 
spin--orbit coupling}.}
\label{TABGER1}
\end{table}

The model Hamiltonian given in equation (\ref{eqGER1}) thus remains a good
approximation for the high concentration limit as was shown for
MnAs. However, there are two important issues when the Mn
concentration is lowered. First, as the splitting between the
minority and majority valence band top becomes smaller the
observed deviations between the model and the density functional
results will become increasingly important. There might be a point
reached where the Kondo form of the $p$-$d$ interaction given in
equation (\ref{eq1}) breaks down. This argument, however, is based 
on a mean field extrapolation into the low density regime. Results 
presented in section IV of this paper indicate that a mean field 
approximation might be invalid for the $p$-$d$ interaction in 
dilute magnetic semiconductors altogether.

Second, as the Mn concentration decreases the band structure, especially around 
the top of the valence band, becomes strongly dominated by 
spin--orbit effects. Hence, scalar--relativistic results in the 
low density limit need to be interpreted very carefully, especially
when spin--dynamics is considered.

\section{G\lowercase{a$_{1-x}$}M\lowercase{n$_x$}A\lowercase{s} in the diluted limit}

The first principle study of Ga$_{1-x}$Mn$_x$ in the low dilution limit is
a formidable theoretical challenge. This is due to the large number of atoms
that one should include in the unit cell in order to reproduce the experimental
Mn concentrations. However several numerical implementations of DFT 
capable of dealing with a large number of atoms are now available. These are generally
based on the use of pseudopotentials \cite{pseudo} and on localized basis sets
\cite{PAO}. Most of the results we will present in the following sections are
obtained with the code {\sc siesta} \cite{Siesta1,Siesta2,Siesta3}, which
combines both of these features.

\subsection{Electronic and magnetic properties}

\begin{figure}[hbtp]
\centerline{\epsfig{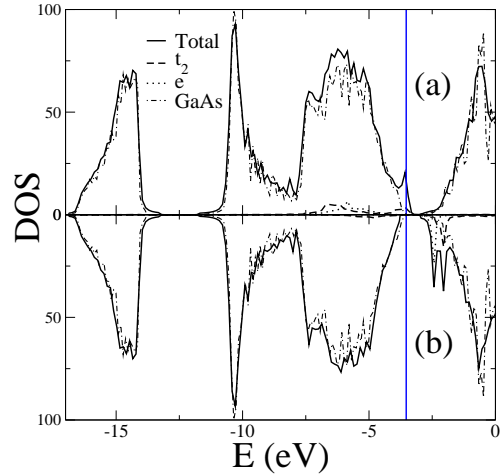}} 
\caption{\small{Partial density of states for Ga$_{1-x}$Mn$_x$As for $x=0.03125$ 
(one Mn$_\mathrm{Ga}$ in a 64 atom GaAs cell): (a)
majority spin, (b) minority spin. The vertical line denotes the position of the
Fermi energy. 
}}
\label{FIG3}
\end{figure}

We start by analyzing the calculated density of states (DOS) of a 64-atom 
GaAs unit cell containing one Mn$_\mathrm{Ga}$ substitution (figure \ref{FIG3}) 
\cite{us2}.
This corresponds to $x$=0.03125, an experimentally accessible concentration. 
From the figure it is clear that Ga$_{1-x}$Mn$_x$As has the electronic structure
of a half-metal. This result is largely confirmed by all density functional
calculations to date both using the LDA \cite{jain,sch,mvs} and the GGA 
\cite{free} approximation. It is also the same structure found for 
In$_{1-x}$Mn$_x$As \cite{akai}. If we project the density of states onto the
different orbital components (partial density of states, PDOS) some additional 
features are revealed.
The majority band exhibits two broad peaks between -4~eV and -1~eV below the 
Fermi energy with strong Mn-$d$ $e$ and $t_2$ component respectively. In addition 
there is a rather narrow $t_2$ peak at the Fermi energy. 
In contrast, the minority band has almost no $d$-character below $E_\mathrm{F}$ 
but instead has two sharp $e$ and $t_2$ peaks around 1~eV above $E_\mathrm{F}$.
The different peak widths reflect the different degrees of hybridization 
of the Mn-$d$ sub-bands with the GaAs bands. 

\begin{figure}[hbtp]
\centerline{\epsfig{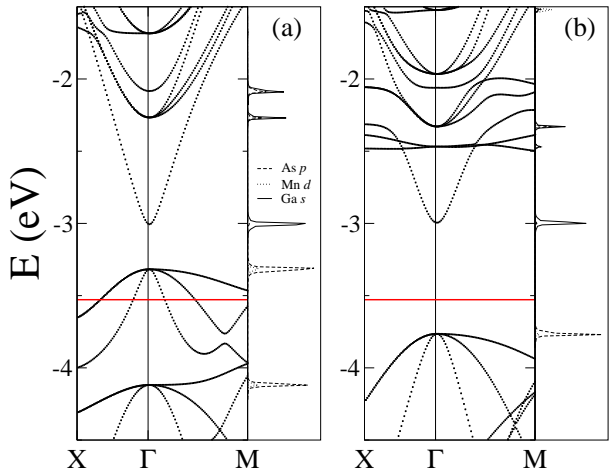}} 
\caption{\small{Band structure and orbital resolved DOS at the $\Gamma$ point
for Ga$_{1-x}$Mn$_x$As with $x$=0.03125 (1 Mn ion in a cubic 64 atom GaAs cell): 
(a) majority band, (b) minority. The horizontal line denotes the position of the
Fermi energy.}}
\label{FIG4}
\end{figure}

The magnetic moment of the unit cell is 4~$\mu_\mathrm{B}$, and remains the same 
up to concentrations of the order of $x$=0.5 \cite{free}. An integer number for the
magnetic moment is consistent with the half-metallicity seen in the DOS. 
At this point it is therefore 
very relevant to discuss the atomic configuration of Mn in GaAs. The band
structure around the $\Gamma$ point for $x$=0.03125 Ga$_{1-x}$Mn$_x$As (1 Mn 
ion in a cubic 64 atom GaAs cell) is presented in figure \ref{FIG4}, along
with the orbital resolved DOS at the $\Gamma$ point.
We consider 
the band structure only along the direction $(\frac{1}{8}\frac{\pi}{c_0},0,0)
\rightarrow(0,0,0)\rightarrow(\frac{1}{8}\frac{\pi}{c_0},
\frac{1}{8}\frac{\pi}{c_0},\frac{1}{8}\frac{\pi}{c_0})$ with $c_0$ the unit 
vector of the cubic cell. We indicate these two directions respectively as 
X and M. 

From the picture it is very clear that the Fermi energy cuts through the top of
the valence band for the majority spin. This, in addition to the fact that all
the Mn-$d$ states in the majority spin band are occupied, suggests that Mn in 
GaAs is incorporated as Mn$^{2+}$ and that there is a polarized hole, which is
antiferromagnetically coupled to the Mn. The presence of a hole is revealed by 
an accurate analysis of the integrated DOS \cite{sch}, and a signature of the
antiferromagnetic coupling is the fact that the induced magnetic 
moment at the As sites is antiparallel to that of the Mn ions \cite{us2,free}. 

It is important to stress that the results from DFT are not as ideal as
the atomic-like picture discussed above, which predicts $S=5/2$ for the Mn. 
Mulliken population analysis \cite{Mul1,Mul2,us2} shows that
the population of the Mn $d$ orbital is 4.7 and 0.7 electronic charges
respectively for the majority and minority states. Considering the fact that the
overlap population is of the order of 0.7, the raw data are compatible with both
Mn $d^4$ and $d^5$. Moreover we must stress that at the top of the valence band
there is a quite large hybridization between Mn $d$ and As $p$ states. All these
factors contribute to reduce the magnetic moment from 5~$\mu_\mathrm{B}$ 
($S=5/2$) expected from the atomic-like picture to 4~$\mu_\mathrm{B}$. A magnetic
moment of 4~$\mu_\mathrm{B}$ appears to be smaller than that found experimentally
($\sim 4.4\mu_\mathrm{B}$) \cite{Ohl1}, although the agreement can be restored by
considering partial hole compensation as we will show in the following sessions.

The behavior of Mn in GaAs seems is similar to that of Mn substituting 
the cation sites in other III-V, and in II-VI and group IV semiconductors. 
Schulthess and Butler \cite{sch} have
calculated the electronic structure of Mn in GaAs, Ge, ZnSe and ZnO. The main
results are that i) the number of minority electrons is not changed by the Mn
impurity and ii) each Mn impurity adds five additional majority states to the
valence band. This leads to a magnetic moment of 3~$\mu_\mathrm{B}$, 
4~$\mu_\mathrm{B}$ and 5~$\mu_\mathrm{B}$, respectively for Mn in Ge, GaAs, and
both ZnSe and ZnO.

It is difficult to extract the localization properties of the holes introduced by 
the Mn ions from DFT calculations. In particular we are not able to conclusively 
establish whether or not the holes are bound to Mn$^{2+}$ forming a neutral (3$d^5+h$) 
complex \cite{Szcz1}. On the one hand plots of the charge density obtained from
states within 0.5~eV around the Fermi energy reveal that most of the charge is
concentrated around the Mn sites \cite{us2}. This seems to suggest localization of the hole
around the Mn ion. On the other hand accurate valence band fitting \cite{jain}
reveals an effective mass quite similar to that of GaAs. This is of course an
indication of delocalization. We believe that this point needs further
investigation. A key element to determining the localization properties of the
holes is a knowledge of the exchange constant $N_0\beta$, which we discuss in the
next Section.

\subsection{The exchange coupling}

We calculate the exchange constant by evaluating the spin-splitting of the
conduction and valence bands. This mimics a typical magneto-optical 
experiment \cite{Szcz2}. The main idea is that in the mean field theory based on
the Hamiltonian of equation (\ref{eq1}) the spin-splitting of the valence band 
depends linearly on both the exchange constant $N_0\beta$ and the Mn concentration 
$x$ \cite{us2}. The same argument holds for the spin-splitting of the
conduction band, which is regulated by a similar Hamiltonian with exchange
constant $N_0\alpha$. Therefore the exchange constants can be directly 
computed from the conduction band-edge (valence band-edge) spin-splittings 
$\Delta E^c=E^c_\downarrow-E^c_\uparrow\;\;$ 
($\Delta E^v=E^v_\downarrow-E^v_\uparrow$) as follows
\begin{equation}
N_0\alpha=\Delta E^c/x\langle S\rangle\:, \;\;\;\;\;\;\; 
N_0\beta=\Delta E^v/x\langle S\rangle
\label{eq3},
\end{equation}
where $\langle S\rangle$ is half of the computed magnetization per Mn ion. 

Recalling the fact that (Ga,Mn)As has a direct gap at the $\Gamma$ point,
we calculate the band structure of Ga$_{1-x}$Mn$_x$As supercells (see
figure \ref{FIG4}) around the $\Gamma$ point for different 
Mn concentrations, and extract the exchange constants by using the equations
(\ref{eq3}). Our results are in table \ref{TAB1}.
\begin{table}[hbtp]
\centerline{
\begin{tabular}{ccccc}
\hline
$x$       &  $\Delta E^c$ (eV) &  $\Delta E^v$ (eV) &  $N_0\alpha$ (eV) & $N_0\beta$ (eV) \\ 
\hline \hline
0.06250   & 0.0339  & -0.6839  &   0.272    &  -5.48    \\ 
0.04166   & 0.0248  & -0.5458  &   0.298    &  -6.54    \\ 
0.03125   & 0.0105  & -0.4472  &   0.168    &  -7.34    \\ 
0.02084   & 0.0099  & -0.3442  &   0.234    &  -8.16    \\ \hline \hline
\end{tabular}}
\caption{\small{Conduction $\Delta E^c$ and valence $\Delta E^v$ band-edge 
spin-splitting, and exchange constants as a function of the Mn concentration 
$x$ for Ga$_{1-x}$Mn$_x$As}.}
\label{TAB1}
\end{table}

The behaviors of the valence and conduction bands are remarkably different. 
For the conduction band, although the spin splitting shows large
fluctuations with $x$, there is no systematic variation with the Mn 
concentration. This indicates that the mean field approximation
that led to equation \ref{eq3} is appropriate
and one can conclude that the exchange coupling between electrons in the 
conduction band and the Mn is ferromagnetic with an exchange constant 
$N_0\alpha\sim$0.2~eV. This is expected since the coupling in this case is
direct (Coulombic $s$-$d$ coupling). Note also that the value of 
the exchange constant $N\alpha$ is very close to that usually found in II-VI 
semiconductors \cite{26}. 

In contrast the valence band shows strong deviation from the mean field
expression (\ref{eq3}), since the valence band spin-splitting does not 
vary linearly with $x$. 
Turning the argument around, $N_0\beta$ increases with decreasing Mn 
concentration, a behavior already well known to occur in Cd$_{1-x}$Mn$_x$S 
\cite{CS1,CS2,CS3}. This suggests that the mean field approximation leading to the
equations (\ref{eq3}) is not appropriate for the valence band of (Ga,Mn)As.

A breakdown of the mean field model occurs when the
potential introduced by the Mn ions is comparable with the relevant band-width. 
We have calculated \cite{us2} the corrections to the mean field model 
by using a free electron model with magnetic impurities described by square 
potentials. The calculation is based on the theory of Benoit \`a la Guillaume,
Scalbert and Dietl, who computed the energy within the Wigner-Seitz approach
\cite{Ben1}.

The main result of the corrected theory is that the mean field approximation tends 
to underestimate the exchange coupling for low dilutions, as observed in our LSDA
calculations. Our estimation of the exchange constant gives a value in the
range of $-4.9~\mathrm{eV}<N_0\beta<-4.4~\mathrm{eV}$, which is very large if
compared with the values quoted by experiments. Moreover it is very
important to note that within this model the valence holes appear to be nearly
bound to the Mn ions. This adds a further indication of the existence of the 
(3$d^5+h$) complex, at least in the low dilution limit.

Finally we want to point out the fact that both the exchange constant and the
valence band spin-splitting are much larger than that found in typical
experiments. There may be several reasons for this disagreement. First, there
is strong experimental evidence in the absence of saturation in the $M-H$
curves at large magnetic fields \cite{Oiw1} and in recent x-ray magnetic dichroism
measurements \cite{Ohl1}, that not all Mn ions contribute to the ferromagnetism. 
Of course an overestimation of $x$ leads to an
underestimation of $N_0\beta$. Secondly the well known lack of accuracy of the
LDA to describe strongly localized charges may result in an over-estimation 
of the $p$-$d$ coupling \cite{zun1}. This of course leads to a larger $N_0\beta$
constant. However, we have shown that LDA does not strongly overbind the NiAs-type
MnAs, and that the structural properties of zincblende MnAs are very similar if 
calculated with LDA or LDA+U. This suggests that the error in the determination
of the $p$-$d$ coupling is not dramatic within LDA.
Therefore we do believe that the two main conclusions from our LDA calculations,
namely that the exchange constant is large enough for the mean field approximation 
to breakdown, and that the holes are nearly bound, are indeed reliable. 
It is interesting to remark that in a
recent paper \cite{SDS} Chattopadhyay, Das Sarma and Millis found a $T_c$ for
(Ga,Mn)As in very good agreement with the experiments, by using 
dynamical mean field theory and our value for the exchange constant.
Moreover the dynamical mean field theory estimation of the critical exchange
constant needed for the formation of an impurity band, agrees very well with 
our value \cite{us2}. This confirms the reliability of our analysis in the 
experimentally relevant regime.

\section{The importance of intrinsic defects}

So far we have always considered the ideal case in which only Mn ions are 
introduced in GaAs. If we assume the nominal valences for Mn (Mn$^{2+}$) 
and Ga (Ga$^{3+}$), we conclude that Mn acts as single acceptor in GaAs.
Therefore an equal concentration of Mn ions and holes is expected. In contrast 
in the actual samples the
hole concentration is much smaller than that of Mn \cite{Ohno1} and
some compensation mechanism occurs. As suggested in the first experimental
works, the presence of As antisites, As$_\mathrm{Ga}$, usually quite abundant in
low-temperature GaAs \cite{spe1}, is likely responsible for the compensation
(As$_\mathrm{Ga}$ in GaAs is a double donor).
Here we summarize the effect of As$_\mathrm{Ga}$ on the magnetic properties 
of (Ga,Mn)As. The detailed results, obtained with {\sc
siesta}, can be found in references \cite{us3,us4,us5}.

\subsection{Compensation due to As$_\mathrm{Ga}$: local effects}

We consider explicitly the effects of the inclusion of As$_\mathrm{Ga}$ in 
(Ga,Mn)As at different dilutions, and study how the chemical environment 
modifies the magnetic interaction between the Mn ions. 
We construct 64 (cubic) and 32 (rectangular) atom GaAs cells in which we 
include two Mn ions (leading to Mn concentrations of respectively $x$=0.0625 
and $x$=0.125) and a variable number of As$_\mathrm{Ga}$ antisites. 
The parameter which quantifies the strength of the exchange coupling is the 
energy difference $\Delta_\mathrm{FA}$ between the total energies of the
antiferromagnetic $E_\mathrm{AF}$ and ferromagnetic $E_\mathrm{FM}$ 
configurations of the supercell. These are obtained by fixing the spin direction
at the beginning of the self-consistent calculation.

We further look at two possible spatial configurations of the Mn ions in the unit
cell: 1) {\it separated}, when the Mn ions occupy positions as far apart as 
possible (i.e. the corner and the middle of the cubic cell), 2) {\it close},
when the Mn atoms occupy two corners of a tetrahedron and are coordinated
through a single As ion (see figure \ref{FIG6}). 

\begin{figure}[ht]
\centerline{\epsfig{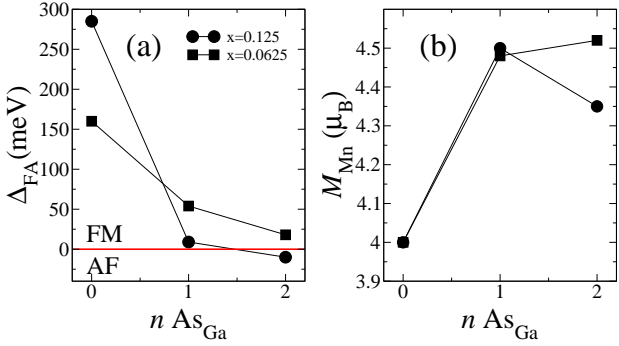}}
\caption{(a) Energy difference between AF and FM alignments, $\Delta_\mathrm{FA}$, 
and (b) magnetization per Mn ion, $M_\mathrm{Mn}$, as a function of the number of 
As$_\mathrm{Ga}$ antisites in the cell: {\it separated} configuration. 
The horizontal line denotes the division between FM and AF alignment.}
\label{FIG5}
\end{figure}

\begin{figure}[ht]
\centerline{\epsfig{file=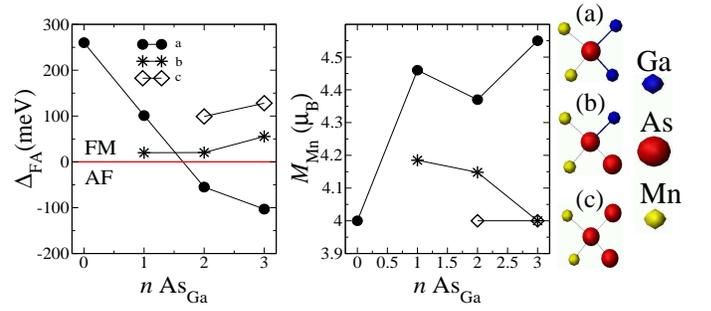,scale=0.45,angle=0}}
\caption{Energy difference between the AF and FM alignments 
$\Delta_\mathrm{FA}$ (left), 
and magnetization per Mn ion $M_\mathrm{Mn}$ (right) as a function of the number of 
As$_\mathrm{Ga}$ antisites in the cell: {\it Close} Mn arrangement. The symbols
$\bullet$, $ * $ and $\diamond$ represent arrangements (a), (b) 
and (c) respectively.}
\label{FIG6}
\end{figure}

$\Delta_\mathrm{FA}$, and the magnetization per Mn ion, $M_\mathrm{Mn}$, for 
the {\it separated} arrangement are presented in figure \ref{FIG5} as a function 
of the number of As$_\mathrm{Ga}$ antisites. The magnetization per Mn ion is 
defined to be half of the magnetization of the cell, calculated in the FM 
aligned phase.
First we note that the ferromagnetic coupling is strongly weakened by
As$_\mathrm{Ga}$ antisite doping. This is expected according to the
picture of hole-mediated ferromagnetism: As$_\mathrm{Ga}$ antisites
contribute electrons into the system and therefore compensate the holes.
We also note that in the case of no antisites, $\Delta_\mathrm{FA}$ is much 
larger in the case of large Mn concentrations (we recall that according to
equation (\ref{eq2}) $T_c$ scales linearly with $x$).
The figure also suggests that the compensation mechanism does not follow the 
nominal atomic valence, since a single As$_\mathrm{Ga}$ antisite per cell is 
not sufficient to destroy the ferromagnetic coupling. 
Above compensation (more that one As$_\mathrm{Ga}$ for two Mn ions) 
antiferromagnetic coupling is obtained for large Mn concentration, while the 
system stays ferromagnetic at low concentration, although in both cases 
$\Delta_\mathrm{FA}$ is rather small ($|\Delta_\mathrm{FA}|\le20$~meV). 
This is consistent with the 
onset of antiferromagnetic super-exchange coupling \cite{paw}, 
the mechanism which is believed to be responsible for the magnetic 
order in the II-VI DMS \cite{26}, at compensation. 
Super-exchange is a short range interaction and therefore is less important in 
the low concentration limit where the Mn ions are well separated. 

These results agree qualitatively with those obtained by Akai \cite{akai} for
(In,Mn)As, using a KKR-CPA-LDA (Korringa-Kohn-Rostoker Coherent Potential
Approximation and Local Density Approximation) method \cite{KKR1}.
Akai interpreted his data as a competition between ferromagnetic
double-exchange and antiferromagnetic super-exchange. 

In figure \ref{FIG5}b we see that $M_\mathrm{Mn}$ increases with the 
As$_\mathrm{Ga}$ concentration and than saturates to a value around 4.5~$\mu_B$. This
can be easily explained by remembering that the top of the majority spin valence band 
of (Ga,Mn)As has some Mn $d$ component due to hybridization (see figure \ref{FIG4}).
As$_\mathrm{Ga}$-doping moves the Fermi energy towards the conduction band, filling
the valence band completely. This enhances the magnetic moment of the unit cell. The
saturation is due to the fact that the next Mn $d$ states available above the 
valence band are at the edge
of the conduction band in the minority spin band. However As$_\mathrm{Ga}$ doping
pins the Fermi energy at mid gap and these states can not be filled. Note that a
magnetic moment per Mn of 4.5~$\mu_B$ is in good agreement with x-ray circular
magnetic dichroism measurements \cite{Ohl1}.

A better insight into the mechanism giving rise to the ferromagnetic order is
given by the results for the {\it close} configuration.
In this case the local microscopic configuration is crucial for the magnetic
properties, therefore we consider three different situations (see figure
\ref{FIG6}):
a) the antisites are far from the Ga$_2$Mn$_2$As$_1$ complex, b) one antisite 
occupies a tetrahedral site (Ga$_1$Mn$_2$As$_2$), and c) two antisites occupy the 
tetrahedral sites (Mn$_2$As$_3$). The results for $x$=0.0625 are presented in
figure \ref{FIG6}.

In case (a) both the $\Delta_\mathrm{FA}$ and $M_\mathrm{Mn}$ curves look 
very similar to those found for the {\it separated} arrangement 
(figure \ref{FIG5}). The main difference is a strong antiferromagnetic coupling
above compensation ($n$ As$_\mathrm{Ga}>$1). Recalling that
Mn in GaAs assumes the $d^5$ configuration ($d$ band half-filled), and that
the Mn $d$ shell is antiferromagnetically coupled with the intermediate As atom 
(through the $p$-$d$ interaction), we conclude that super-exchange 
coupling stabilizes the AF phase at and above compensation in the Ga$_2$Mn$_2$As$_1$ 
complex. Note that the same conclusions have been drawn by Park {\it et al.} for
Ge$_{1-x}$Mn$_x$ \cite{park1}, although in that case the antiferromagnetic
coupling seems to occurs also if holes are present.

Cases (b) and (c) present several interesting features. The most remarkable
is that the ferromagnetic alignment is stable and almost insensitive to 
the total As$_\mathrm{Ga}$ concentration. This suggests that the dominant
interaction in the Ga$_1$Mn$_2$As$_2$ and Mn$_2$As$_3$ complexes is completely
local. Once again Mulliken population analyses help in understanding the electronic
configuration of these complexes. 
\begin{table}[hbtp]
\begin{tabular}{cccccccc}
\hline
$n$~As$_\mathrm{Ga}$ & Type & Mn-$d_\uparrow$ & Mn-$d_\downarrow$ & As-$p_\uparrow$ &
As-$p_\downarrow$ & As-$p$ & As \\ \hline \hline
2 & a  & 4.74 & 0.70 & 1.55 & 1.63 & 3.18 & 4.92 \\ 
2 & b  & 4.74 & 0.71 & 1.50 & 1.67 & 3.17 & 4.95 \\ 
2 & c  & 4.72 & 0.75 & 1.44 & 1.71 & 3.15 & 4.97 \\ \hline \hline

3 & a  & 4.76 & 0.68 & 1.57 & 1.63 & 3.20 & 4.93 \\ 
3 & b  & 4.74 & 0.74 & 1.55 & 1.65 & 3.20 & 4.97 \\ 
3 & c  & 4.73 & 0.75 & 1.49 & 1.69 & 3.17 & 4.99 \\ \hline \hline
\end{tabular}
\caption{Mulliken atomic and orbital populations for the Mn ions and the
intermediate As atom of the complexes of figure \ref{FIG6}. The Mn concentration 
is $x$=0.0625. 
The symbols $\uparrow$ and $\downarrow$ correspond to majority and minority 
spin respectively. The populations are in units of the electronic charge $|e|$.}
\label{TAB2}
\end{table}
In table \ref{TAB2} we present the Mulliken orbital population for the two 
Mn ions and the intermediate As ion of the complexes of Fig.\ref{FIG6}. 
To understand the table, imagine the situation in which 
two As$_\mathrm{Ga}$ antisites are first located far from the Mn-As-Mn complex, 
and then each antisite is moved in turn to one of the two other corners of the 
tetrahedron. This corresponds to moving vertically (top to bottom) 
down the first half of the table.
We notice that: i) the charge on the middle As atom increases, ii) the 
spin-polarization of the $p$ shell of the middle As atom increases, and iii) 
the total population of the $p$-shell of the middle As atom decreases. 
And most importantly the magnetic coupling changes from antiferromagnetic to 
ferromagnetic.

It is well-known that in an antiferromagnetic crystal the presence of a bound 
carrier (electron or hole) which is Zener coupled to the local spins 
{\it always} induces a distortion in the antiferromagnetic
lattice \cite{DG}. We therefore propose that the observed transition from
antiferromagnetic in Ga$_2$Mn$_2$As$_1$ to ferromagnetic coupling in Ga$_1$Mn$_2$As$_2$ and
Mn$_2$As$_3$ results from the onset of ferromagnetic double-exchange
coupling mediated by a bound Zener carrier. It is important to note that
the potential necessary to bound an extra charge in the vicinity of the Mn ions
is provided by the As antisites. We also emphasize that if Ga$_2$Mn$_2$As$_1$ 
or Mn$_2$As$_3$ complexes are present in actual samples, then the Mn ions 
in the complexes will not contribute to the overall ferromagnetic 
alignment, being magnetically ``locked'' by the local
environment. This is a possible explanation of the fact that in actual
samples only a fraction of the Mn ions contribute to the ferromagnetism 
\cite{Oiw1,Ohl1}. Similar conclusions based on DFT calculations have 
been drawn for Ge$_{1-x}$Mn$_x$ \cite{park1}.

\subsection{Defect manipulation}

\begin{figure}[ht]
\centerline{\epsfig{file=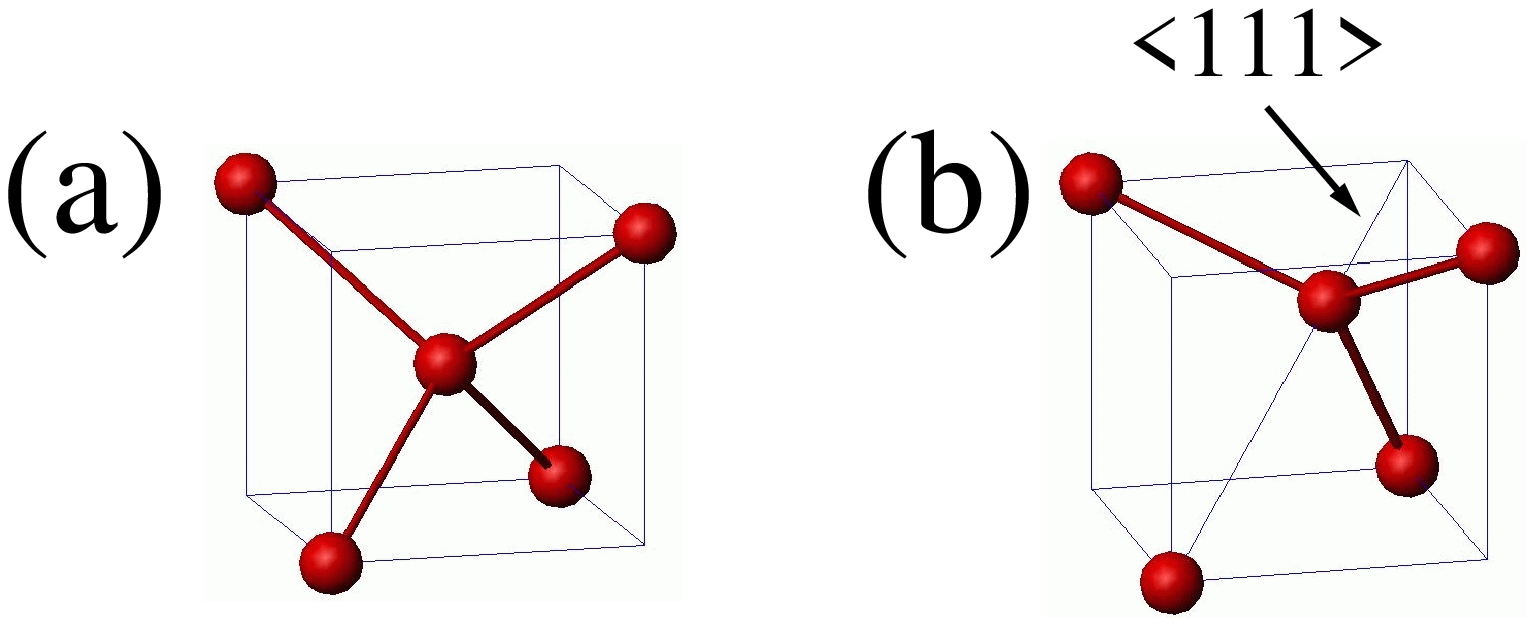,scale=0.4,angle=0}}
\caption{(a) As$_\mathrm{Ga}$ and (b) As$_i$-V$_\mathrm{Ga}$ pair obtained by moving 
As$_\mathrm{Ga}$ along the $\langle111\rangle$ direction.}
\label{FIG7}
\end{figure}

In a recent paper \cite{us4} we have shown that intrinsic defect manipulation
can provide a valuable way of tuning the carrier concentration and hence enhancing 
the $T_c$ in (Ga,Mn)As, without changing the Mn concentration nor the microscopic
configuration of the Mn ions. Here we summarize the basic ideas. We first recall 
that isolated As$_\mathrm{Ga}$ in GaAs are responsible for the photoquenchable 
EL2 defect \cite{Dab}. In fact upon illumination As$_\mathrm{Ga}$ 
undergoes a structural transition to an As interstitial-Ga vacancy
(As$_i$-V$_\mathrm{Ga}$) pair, which is obtained by moving As$_\mathrm{Ga}$ 
along the $\langle111\rangle$ direction (see figure \ref{FIG7}). 
This complex is metastable since As$_\mathrm{Ga}$ can be regenerated by heating. 
It is crucial to observe that the As$_i$-V$_\mathrm{Ga}$ pair is not electronically 
active in GaAs, since its only state in the bandgap is completely filled.
In what follows we show that this metastable complex is present, can be obtained 
by illumination and is also electronically inactive in (Ga,Mn)As. 

\begin{figure}[ht]
\centerline{\epsfig{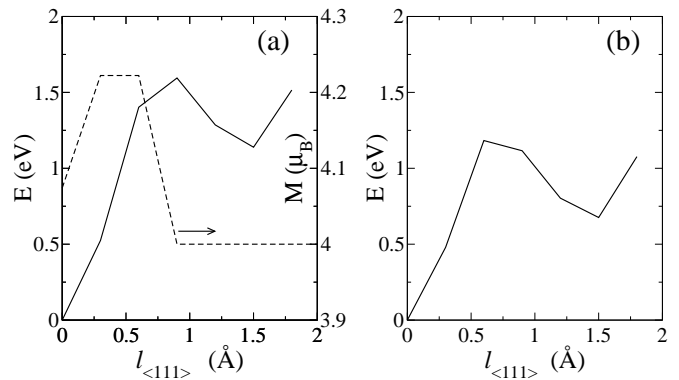}}
\caption{Total energy (left-hand side scale) and magnetization (right-hand side 
scale) for (a) (Ga,Mn)As and (b) GaAs as a function of the displacement of an 
As$_\mathrm{Ga}$ antisite along $\langle 111 \rangle$. The energy of 
$l_{\langle 111 \rangle}$=0 is set to 0~eV.}
\label{FIG8}
\end{figure}

In figure \ref{FIG8} we present the total energy and the magnetization 
of a 64 atom unit cell containing one Mn ion ($x$=0.03125) and one 
As$_\mathrm{Ga}$ (note that in this case the system is $n$-doped, with one excess
electron per unit cell) as a function 
of the displacement $l_{\langle111\rangle}$ of As$_\mathrm{Ga}$ along 
$\langle111\rangle$. 
For comparison we also present the same curve for GaAs, which is in very
good agreement with previously published results \cite{Dab}. It is clear that also
in (Ga,Mn)As the As$_i$-V$_\mathrm{Ga}$ defect is metastable and that the 
energy barrier for the thermal regeneration is 0.45~eV. This, according to
kinetic calculations \cite{Dab} gives a regeneration temperature of about
100~$K$. 

Having established that the As$_i$-V$_\mathrm{Ga}$ defect is metastable
let us now prove that it can be generated also in (Ga,Mn)As. The mechanism has been
explained by Scheffler {\it et al.} \cite{Dab,Sch1}, and can be summarized as
follows.
It is known that the excited $a_1^1t_2^1$ electronic configuration of a 
tetrahedral substitutional double donor induces lattice 
distortion. Such distortion occurs is As$_\mathrm{Ga}$, which possesses a doubly
occupied donor level $a$ at midgap and an empty resonant state  with $t_2$ 
symmetry close to the conduction band edge.
The distortion is initiated because the many-electron wave function of the 
$a_1^1t_2^1$ configuration is orbitally degenerate therefore the system is 
Jahn-Teller unstable. Jahn-Teller distortion splits the $t_2$ state into a lower $a$ state 
(half-filled) and a higher $e$ state (empty). Therefore an optical excitation of 
the $a_1^2t_2^0$ ground-state to the $a_1^1t_2^1$ will initiate a distortion. 
It has been demonstrated that in GaAs the total-energy 
curve as a function of the displacement of As$_\mathrm{Ga}$ along 
$\langle111\rangle$ for the $a_1^1t_2^1$ configuration has a minimum for 
$l_{\langle111\rangle}\sim$0.3\AA\ \cite{Dab}. Then the system has some probability 
of relaxing onto the 
As$_i$-V$_\mathrm{Ga}$ side of the total-energy curve of figure \ref{FIG8}, ending
up considerably far from the antisite position, creating the 
As$_i$-V$_\mathrm{Ga}$ pair.

The same mechanism holds for (Ga,Mn)As since the only difference with respect to
GaAs is the spin-splitting of the bands. This splitting however does not
move the $a$ state of As$_\mathrm{Ga}$ into the valence band nor the $t_2$ state 
into the conduction band. With these considerations it is clear that 
the As antisite in (Ga,Mn)As presents the same features as that in GaAs and, 
therefore, the mechanism described above is still applicable.

\begin{figure}[ht]
\centerline{\epsfig{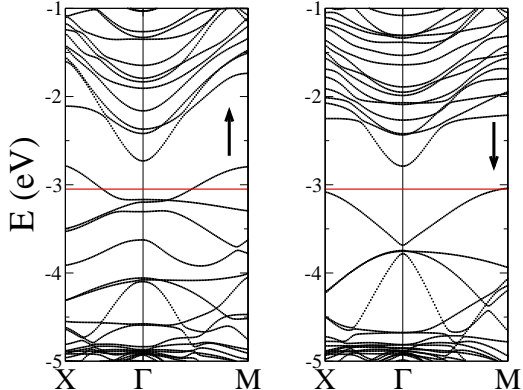}}
\caption{Band structure for Ga$_{1-x}$Mn$_x$As with intrinsic defects:
As$_\mathrm{Ga}$. These are the bands for a cubic 64 atom supercell
(2$\times$2$\times$2 zincblende cubic cells) containing one Mn ion and one
As$_\mathrm{Ga}$. On the right-hand side the majority spin and on the left the minority.
The horizontal line denotes the position of the Fermi energy.}
\label{FIG9}
\end{figure}

\begin{figure}[ht]
\centerline{\epsfig{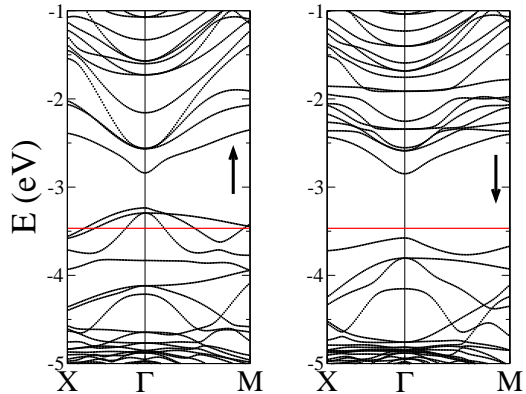}}
\caption{Band structure for Ga$_{1-x}$Mn$_x$As with intrinsic defects:
As$_i$-V$_\mathrm{Ga}$ pair. These are the bands for a cubic 64 atom supercell
(2$\times$2$\times$2 zincblende cubic cells) containing one Mn ion and one
As$_i$-V$_\mathrm{Ga}$ pair. On the right-hand side the majority spin and on the left the minority.
The horizontal line denotes the position of the Fermi energy.}
\label{FIG10}
\end{figure}

The effects of this transition on the ferromagnetic properties are quite
dramatic. We calculate the energy difference between the ferromagnetic and
antiferromagnetic alignment, $\Delta_\mathrm{FA}$, and we observe an increase from
53~meV to 124~meV when a single As$_\mathrm{Ga}$ is transformed into a 
As$_i$-V$_\mathrm{Ga}$ pair. Hence the ferromagnetic order is
strengthened. The reason for this enhancement of the ferromagnetic coupling is
that most of the hole compensation introduced by As$_\mathrm{Ga}$ is lifted by the
creation of the As$_i$-V$_\mathrm{Ga}$ pair. The band structures of (Ga,Mn)As
containing As$_\mathrm{Ga}$ and As$_i$-V$_\mathrm{Ga}$ pair shown in figures
\ref{FIG9} and \ref{FIG10} explain this point.
It is clear that in the case of As$_\mathrm{Ga}$ the Fermi energy cuts through 
the As$_\mathrm{Ga}$ impurity band, leaving the GaAs valence band completely 
filled (note that the donor level is only marginally spin-split, although it is
hybridized with the top of the valence band). This means
that the compensation is complete (note that in the figure we have one Mn ion 
and one As$_\mathrm{Ga}$ in the cell, therefore we correctly expect an excess of 
one electron in the impurity band). 
In contrast in the case of the As$_i$-V$_\mathrm{Ga}$ pair
there are still holes in the system, since the Fermi energy cuts through 
the top of the valence band and the impurity level. It is also important to note
that the impurity level has small hybridization with the majority valence band 
of (Ga,Mn)As. This means that in this case there are holes in the system with a
dispersion very close to that of the defect free case. For this reason the
presence of As$_i$-V$_\mathrm{Ga}$ pairs does not alter the magnetic properties
compared with the defect-free alloy (note that $\Delta_\mathrm{FA}$ for this 
situation is 124~meV, very similar to the defect free value of 159~meV).

In conclusion, the optically induced transition of As$_\mathrm{Ga}$ to
a As$_i$-V$_\mathrm{Ga}$ pair reduces the hole compensation, strengthening the
ferromagnetic coupling. This, in principle, allows tuning of the hole concentration
without acting on the chemical composition of a sample. However, since the 
antisites regenerate for temperatures of the order of 100K, the 
mechanism cannot be used to obtain very high $T_c$s.

\section{Diluted Ferromagnetic Heterostructures: magnetic and transport 
properties}

We have seen in the previous Sections that formation of the NiAs-type phase
of MnAs is easier than formation of the zincblende phase, which 
in turn is very desirable due to its half-metallic band structure.
It has been demonstrated \cite{Kaw1}, however, that zincblende MnAs can be grown up to
half of a monolayer in GaAs/MnAs superlattices. These structures are called
digital ferromagnetic heterostructures (DFH). Since our earlier work has shown
a strong dependence of the magnetic properties on the local arrangement of the Mn
ions, we expect that the properties of DFHs would be very different from their 
random alloy counterparts.

These are the main experimental findings \cite{Kaw1}: first, $T_c$ decays with
increasing GaAs layer thickness separating the MnAs sub-monolayers, and 
saturates for thicknesses larger than $\sim$50 GaAs monolayers. 
The saturation is unexpected according to the mean field model for three dimensional 
systems, since the total Mn concentration in the sample decreases with the 
increase of the GaAs thickness. 
This separation dependence suggests that DFH's behave like planar systems.

Secondly, Hall measurements in the direction parallel to the MnAs planes show 
an anomalous Hall effect for undoped samples, which disappears upon Be-doping 
\cite{gwin1,gwin2}. Large Shubnikov de Haas oscillations are found in 
doped samples, although surprisingly the charge densities
extracted from the Hall coefficient and from the Shubnikov de Haas oscillations
are different. This suggests that two different carrier types could
be present in the system. 

Density functional theory has answered the following questions regarding the 
physics of DFHs:
i) what is the real dimensionality of the system? ii) are the carriers 
spin-polarized? iii) what is the carrier distribution in the system?

\begin{figure}[hbtp]
\centerline{\epsfig{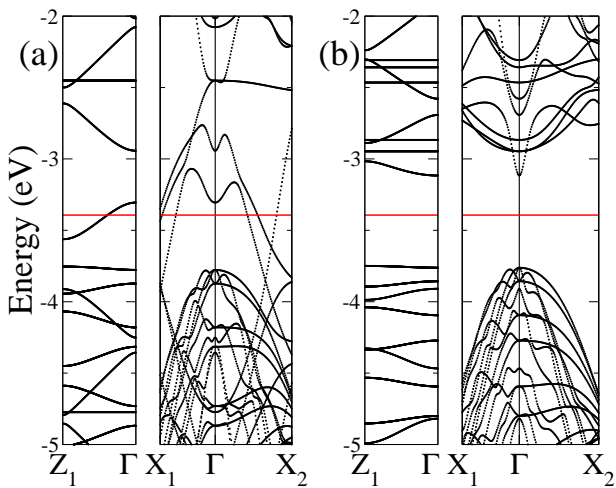}}
\caption{Band structure for a MnAs$_1$/GaAs$_{15}$ superlattice: (a) majority
and (b) minority spins. $X_1$ and $X_2$ are in the MnAs plane and
denote directions along respectively the edge and the diagonal of the cubic
supercell. $\Gamma\rightarrow Z_1$ is the direction orthogonal to the MnAs
plane. The horizontal line indicates the Fermi energy.}
\label{FIG11}
\end{figure}

We have used the code {\sc siesta} \cite{us6} with a  
DFH superlattice constructed from $N$ GaAs cubic cells (8 atoms in the cell) aligned
along the $z$ direction. One Ga plane (two atoms) is substituted 
with Mn and periodic boundary conditions are applied. 
This leads to an infinite MnAs$_1$/GaAs$_{2N-1}$ superlattice, where MnAs 
zincblende monolayers are separated by a $5.65\times N$\AA\ thick GaAs layer.
In figure \ref{FIG11} we present the calculated band structure for the case $N=8$ (with a 
45.2\AA\ thick GaAs inter-layer), for both the majority and minority spins.
It shows clearly a very peculiar half-metallic structure. Although the band
structure has a gap for the minority spin and some band crossing at the
Fermi energy for the majority, the band dispersion in the MnAs plane 
($X_1\rightarrow\Gamma\rightarrow X_2$) is quite broad while the one
perpendicular to the MnAs plane ($Z_1\rightarrow\Gamma$) is very narrow
(impurity-like band). Therefore MnAs/GaAs looks like a {\it two dimensional} 
half metal with small hopping between the MnAs planes. 

We also look at the stability of the ferromagnetic phase as a function 
of the separation between Mn planes, by calculating $\Delta_\mathrm{FA}$.
Surprisingly $\Delta_\mathrm{FA}$ is independent of the GaAs thickness for 
the range of thicknesses investigated here (531~meV, 533~meV and 515~meV 
respectively for $N$=4, $N$=6 and $N$=8). This is consistent with the
experimental insensitivity of $T_c$ to the GaAs thickness. This first analysis
shows that most of the physics of DFH occurs in the MnAs planes. 

In order to have a better understanding we have also performed transport 
calculations in the ballistic limit.
One of the advantages of using a localized basis set for the DFT calculation is 
that, at the end of the self-consistent procedure a tight-binding Hamiltonian
is generated by direct numerical integration over a 
real space grid \cite{Siesta3}. Then transport properties can be calculated 
in the ballistic limit by using a well established technique for tight-binding 
Hamiltonians, which is described in reference \cite{tran}. For this work, we 
have generalized the technique to the case of non-orthogonal tight-binding models 
with singular coupling matrices \cite{us6}. 
More details can be found in the cited literature.

\begin{figure}[hbtp]
\centerline{\epsfig{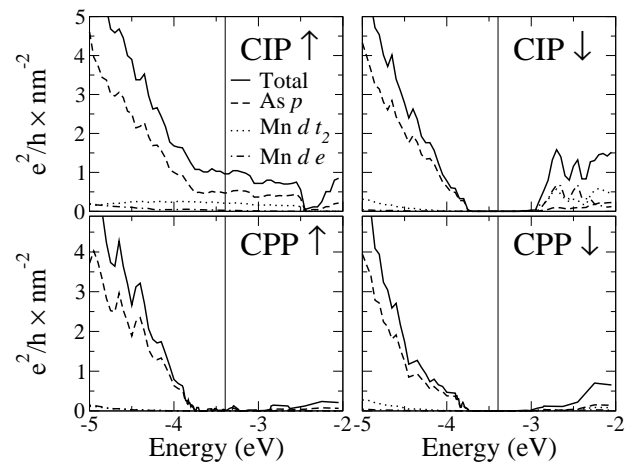}}
\caption{Total and partial conductance per unit area for a MnAs$_1$/GaAs$_{15}$ 
superlattice as a function of the position of the Fermi energy for 
majority ($\uparrow$) and minority ($\downarrow$) spin bands. 
The vertical line denotes the position of the Fermi energy for undoped samples.}
\label{FIG12}
\end{figure}

In figure \ref{FIG12} we present our calculated conductance per unit area as a 
function of the position of the Fermi energy for a MnAs$_1$/GaAs$_{15}$ 
superlattice for both the current in the MnAs plane (CIP) and current 
perpendicular to the Mn plane (CPP) directions, and for both spins. 
We also project the conductance onto the atomic orbital basis set in order to 
determine the orbital character of the electrons carrying the current \cite{tran}.
As expected from the band structure, the ballistic current is 100\% 
spin polarized, with no current for the minority spin band. In the majority spin
band the behavior is very different for the CIP and CPP alignment. In the CIP
case the conductance is quite large and independent of the energy, with a 
significant contribution (roughly 20\% of the total conductance) coming from 
the Mn $d$ $t_2$ orbitals. In contrast, the conductance is very small 
in the CPP direction with orbital contribution almost entirely from the As $p$ states.
Moreover the conductance at $E_\mathrm{F}$ comes from just a few $k$-points around 
the $\Gamma$ point (in the direction orthogonal to the transport). These 
correspond to the states with the largest kinetic energy in the direction of 
the transport. This situation is similar to that occurring in tunneling junctions and 
so we describe the transport as tunneling-like, meaning that the transport in
the CPP direction is through hopping between the MnAs planes.

\begin{figure}[hbtp]
\centerline{\epsfig{file=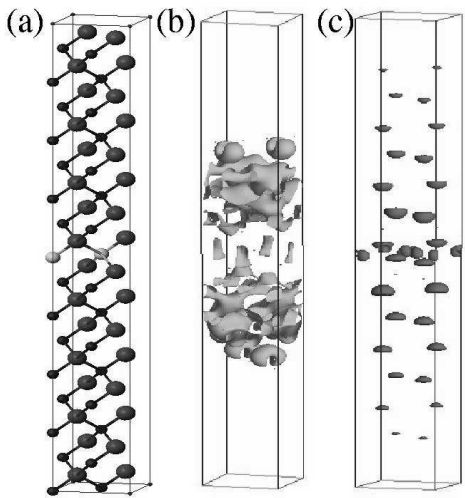,scale=0.3,angle=0}}
\caption{Charge density distribution in real space for MnAs$_1$/GaAs$_{15}$ (a)
calculated only for those states contributing to the conductance and 
energy within 0.3~eV from $E_\mathrm{F}$: CIP (b) and CPP (c) configurations.}
\label{FIG13}
\end{figure}

Finally we investigate the spatial distribution of the current. This is given
by the charge density distribution in real space, $\rho({\bf r})$, calculated 
only for those states contributing to the conductance (see figure \ref{FIG13}) 
\cite{us6}. 
The figure confirms that the current in the CIP case is distributed mainly in
a narrow region around the MnAs planes, with small spillage outside.
In contrast, the CPP current is mainly located at the Mn plane with small 
contributions from the GaAs layers. This means that carriers are strongly 
confined in the MnAs plane and the perpendicular transport is via hopping 
between the planes.

DFT has therefore answered all the questions stated at the beginning of
this Section.
The DFHs appear to be two dimensional half-metals, with a metallic-like ballistic
conductance in the MnAs plane and an hopping conductance perpendicular to the
MnAs planes. The ferromagnetism is therefore insensitive to the GaAs layer 
thickness, showing no dilution effect.

\section{Other diluted magnetic semiconductors}

So far we have concentrated primarily on the properties of (Ga,Mn)As, which
is the most well studied of the DMSs. However, in the last year several other DMSs 
have been synthesized, some of them showing remarkably high $T_c$. At the same time several
DFT calculations have been published both explaining the properties of existing 
materials, or making predictions for new ones. The prediction of material
properties in advance of experiments is one of the most important aspects of 
density functional theory. Here we review the principal works in this area.

\subsection{Ga$_{1-x}$M$_x$As with M=V, Cr, Fe}

Doping of GaAs with transition metals other than Mn has been investigated
theoretically using both the full potential linearized augmented plane wave (FLAPW) 
method \cite{shirai1} and the atomic sphere approximation \cite{mvs}. Since the
first method is computationally more demanding the author concentrates on the
electronic structure of hypothetical zincblende MAs with M=V, Cr, Mn, Fe,
and on the high concentration limit of Ga$_{1-x}$M$_x$As ($x$=0.25 and $x$=0.125). 
The main result is that, while the ground state of FeAs is antiferromagnetic, 
VAs, CrAs, and MnAs at the respective equilibrium lattice constants appear to be 
half-metallic ferromagnets with magnetic moments respectively of 2~$\mu_\mathrm{B}$, 
3~$\mu_\mathrm{B}$ and 4~$\mu_\mathrm{B}$ per formula unit. 

Turning attention to Ga$_{1-x}$M$_x$As it is interesting to note that
the calculated DOS for Ga$_{1-x}$V$_x$As in the large concentration limit reveals 
very small hybridization of the V 3$d$ orbitals with the
GaAs valence band, suggesting that in this compound the magnetic coupling may
be rather small. In contrast, Ga$_{1-x}$Cr$_x$As shows a DOS very similar to that 
of Ga$_{1-x}$Mn$_x$As. 
This suggests that Ga$_{1-x}$Cr$_x$As may be a potential
candidate for a high $T_c$ DMS. In addition 
van~Schilfgaarde and Mryasov fit their DFT results with a pairwise
Heisenberg-like energy $E=-\sum_{ij}J_{ij}\vec{s}_i\cdot\vec{s}_j$, and found
that the values of $J$ for Ga$_{1-x}$Cr$_x$As are larger than those of 
Ga$_{1-x}$Mn$_x$As. This suggests stronger ferromagnetism for Ga$_{1-x}$Cr$_x$As
than for Ga$_{1-x}$Mn$_x$As. However this prediction is strongly affected by
the fact that the DFT results strongly deviate from the Heisenberg-like form.

Experimentally both Ga$_{1-x}$Fe$_x$As \cite{FeAs} and Ga$_{1-x}$Cr$_x$As
\cite{CrAs} have been synthesized and there is a little evidence for ferromagnetic 
order.
However very recently zincblende CrAs has been successfully grown on GaAs at low
temperature \cite{CrAs2}, showing a $T_c$ above room temperature. Moreover
preliminary measurements show a magnetic moment per formula unit very close to
3~$\mu_\mathrm{B}$, in good agreement with the theoretical predictions \cite{shirai1}.
To our knowledge Ga$_{1-x}$V$_x$As has never been grown.

\subsection{Ga$_{1-x}$Mn$_x$N}

There has been a large effort in synthesizing Ga$_{1-x}$Mn$_x$N in the last two
years, since Dietl {\it et al.} predicted a Curie temperature higher than any other 
Mn-doped semiconductors for this compound \cite{Dietl1}. 
Very recently this prediction has been confirmed 
experimentally \cite{GaMnN} with Ga$_{1-x}$Mn$_x$N
at 10\% Mn concentration showing a remarkably large $T_c$ (940~K).
Unfortunately the mechanism for the ferromagnetism in Ga$_{1-x}$Mn$_x$N
is not clearly understood due to the lack of a experimental data. In particular
the nature and the role of the free carriers (if present) are not yet clear.

Sato and Katayama-Yoshida performed DFT calculations within the
Korringa-Kohn-Rostoker method and the coherent potential approximation
(KKR-CPA) \cite{sato}. They looked at transition metal doping of GaN and studied
the stability of the ferromagnetic configuration against the spin-glass state.
The ferromagnetic state appears to be stable for (Ga,V)N and (Ga,Cr)N, while for 
(Ga,Fe)N, (Ga,Co)N and (Ga,Ni)N the spin glass state has lower total energy
at every concentrations studied ($x$=0.05, 0.1, 0.15, 0.20, 0.25).
The case of (Ga,Mn)N is critical since the ferromagnetic state is
stable only at low Mn concentration ($x<$0.15). This is explained in terms of
competition between double- and super-exchange, with the latter being dominant
for small Mn-Mn separations. 

The DOS of (Ga,Mn)N looks rather different that of (Ga,Mn)As, 
showing a much stronger $d$
contribution at the Fermi energy and smaller $p$-$d$ hybridization. 
This leads to the formation of a Mn $d$ impurity band in the band gap
of GaN, as calculated by Kronik {\it et al.} \cite{kron}, who also calculated that
the (Ga,Mn)N valence band is not spin-split.
The atomic configuration of Mn is calculated to be Mn$^{3+}$ with the $d$ orbitals
arranging as $d^4$ \cite{sato}.
This contradicts the experimental results for paramagnetic 
(Ga,Mn)As \cite{zajac}, which convincingly show a $d^5$ ($S=5/2$) configuration.
The reasons of this discrepancy are not clear at the moment. 

\subsection{Zn$_{1-x}$M$_x$O with M=V, Cr, Mn, Fe, Co, Ni}

The mean field model of Dietl {\it et al.} also predicts a very large $T_c$ 
for Mn-doped ZnO \cite{Dietl1}, provided that the sample is $p$-doped. 
Sato and Katayama-Yoshida also investigated the stability of the 
ferromagnetic phase with respect to the
spin-glass phase for Zn$_{1-x}$M$_x$O (M=V, Cr, Mn, Fe, Co, Ni), again using a
KKR-CPA method \cite{sky1,sky2,sky3}. They found a ferromagnetic ground state
for all the materials except Mn, for which the spin-glass configuration has a lower
energy. However they also showed that in the case of Mn the ferromagnetic
configuration can be obtained by large $p$-doping (they substitute N atoms at
the O sites in their calculations). 

Since in the II-VI semiconductors only very small hole concentrations can 
be obtained by doping, they also investigated the conditions to have
strong ferromagnetism in transition-metal-doped ZnO with additional $n$ doping
(Ga at the Zn sites). The result of this calculation is that the presence of electrons stabilizes the
ferromagnetic order in ZnO doped with Fe, Co and Ni \cite{sky2}. It is important
to note that in all these cases the Mn $d$ shell is more than half-filled and
that the Fermi energy is located within a Mn $d$ region of the minority
spin density of states. These are the conditions for strong double-exchange
coupling between the transition metal ions. The fact that $n$-doped (Zn,Co)O 
\cite{ZnCoO} and (Zn,Ni)O \cite{ZnNiO} have been grown, showing $T_c$s above 
room temperature, is very encouraging.

Similar calculations have been carried out for transition-metal-doped ZnS,
ZnSe and ZnTe \cite{sky4,shoren} without any additional doping. These show that
only V- and Cr- doped materials have a ferromagnetic ground state, which is due
to double-exchange coupling. In this case the carriers mediating the
double-exchange are holes at the top of the valence band, which in turn are
strongly hybridized with the $p$ states of the group VI element. It is worth
mentioning that to date no ferromagnetic order has been found experimentally 
in these materials. 
However real samples present strong self-compensation, which
suppresses the hole-mediated double-exchange mechanism.

\subsection{M$_{1-x}$Mn$_x$GeII$_2$ with M=Cd, Zn and II=P, As}

Cd$_{1-x}$Mn$_x$GeP$_2$ was the first room temperature diluted ferromagnetic
semiconductors to be grown \cite{CdMnGeP}. It has a body-centered tetragonal cell 
and the main advantage of this structure is that Mn can substitute for the II 
cations, adopting the Mn$^{2+}$ state (``natural'' for Mn). 
Accurate calculations both with GGA and LDA have been performed 
\cite{af1,af2}, showing that indeed Mn assumes a 
Mn$^{2+}$ state. The calculations demonstrate a very weak sensitivity of the
magnetic and electronic structure on both the anion (As or P), and cation (Cd or
Zn) elements. 

Interestingly both the LDA and GGA results give an
antiferromagnetic ground state, contradicting the experimental results.
However two aspects must be pointed out. First, the samples in the experiments 
\cite{CdMnGeP} are prepared with vacuum deposition on a single crystal surface,
followed by solid state reaction at high temperature. This means that most of 
the Mn ions are located close to the surface. The DFT calculations are performed
for perfect crystalline bulk systems, whose properties will certainly differ
from those of the actual samples. Secondly it has been demonstrated \cite{af2} 
that the ferromagnetic phase can become stable against the antiferromagnetic
one, if electrons are introduced in the system (S substituting for P). This
seems to suggest that, the presence of free carriers is also essential for 
ferromagnetic order in chalcopyrite semiconductors.

Finally, very recently Mahadevan and Zunger calculated the formation 
energies of several intrinsic defects of CdGeP$_2$, of Mn impurities in CdGeP$_2$ 
either at the Cd and the Ge site, and of complexes of these \cite{mazu}. They 
found that under Cd, Mn and P rich growth conditions Mn$_\mathrm{Ge}$ impurities 
naturally form in the crystal. 
These impurities, that in contrast to Mn in GaAs do not form clusters, 
are acceptors in CdGeP$_2$ and align ferromagnetically. In addition it is important 
to observe that the ferromagnetic coupling is also found between 
Mn$_\mathrm{Ge}$ and Mn$_\mathrm{Cd}$ impurities. These predictions 
offer very good guidelines for growing room-temperature Mn-doped CdGeP$_2$.
    
\section{Conclusions and future directions}

In this paper we have presented the recent DFT contributions to the physics 
of diluted magnetic semiconductors. Most of the results are for (Ga,Mn)As, 
have been obtained within either the LSDA or the GGA approximation and
capture most of the relevant physics of (Ga,Mn)As. 
We have also summarize the latest results for other DMSs, whose
properties are less well studied and understood. 

During these studies several disagreements with experiments have appeared, 
suggesting that in some cases the LSDA/GGA approximations 
may be not completely appropriate. It
is therefore indispensable to understand the limit of the reliability of the 
LSDA/GGA approximations and eventually to correct them. It is well known that
the local density approximation fails in describing systems where
electron-electron interaction is strong. This generally happens when we try to 
describe orbitals tightly bound to their nuclei. The $d$ orbitals of 
transition metals in transition-metal oxides are an example of this.

A typical signature of the inappropriateness of the LSDA/GGA is that the bands from
tightly bound orbitals are pushed towards higher energies. This can create
the following problems in DMSs: 1) underestimation of the true band occupation, leading
to an erroneous magnetic moment, 2) overestimation of $p$-$d$ interaction, due
to artificially strong hybridization, 3) overbinding of the structure, 4)
tendency to metallicity in otherwise semiconducting systems, 5) prediction of
the wrong magnetic state (ie ferromagnetic instead of antiferromagnetic). All
these problems can of course strongly affect the predictions of DFT calculations,
and they must be corrected. 

Several schemes have been adopted to go beyond the LSDA/GGA approximation, and
here we briefly list them. In 1981 Perdew and Zunger \cite{pzu} showed that most
of the problems of the LDA approximation come from its incorrect treatment of 
coulombic self-interaction. They demonstrated that self-interaction corrections 
(SIC), namely the subtraction of spurious self-interaction from the density functional,
greatly improve the description of localized states. Since then several
implementations have been made \cite{svane} including algorithms using 
pseudopotentials \cite{vogel}.

A quite different approach to this problem is represented by the LDA+U scheme
\cite{anis1,anis2}. The main idea is to combine density functional theory with 
an Hubbard description of those orbitals which suffer strong correlations.
In this case an additional Hubbard-like term is added to the energy functional 
and the consequent single particle equations are deduced. The theory gain two
extra parameters (the Hubbard parameter $U$ and the Hund's rule exchange $J$),
which can be fixed from experimental values or calculated self-consistently. 
The advantages of the LDA+U scheme are the relatively easy numerical implementation 
and the fact that the parameters can be directly compared with those coming from 
simpler models.

In conclusion, we have shown that DFT can be a very powerful tool for describing
structural, electronic, magnetic and transport properties of diluted magnetic
semiconductors. In particular DFT is very useful for obtaining parameters not
accessible from experiments, and for providing accurate descriptions of properties
occurring at the atomic scale. The prediction of new materials with novel
properties is possible, although special care should be taken in order to keep
errors coming from the local density approximation under control.

\section*{Acknowledgments}

We would like to thank S.~Picozzi, L.~Kronik, S.~Hellberg, P.~Mahadevan
and B.I.~Min for useful discussions, comments and permission to discuss unpublished 
results.
This work made use of MRL Central Facilities supported by the National Science 
Foundation under award No. DMR96-32716. This work is supported by ONR 
grant N00014-00-10557, by NSF-DMR under the grant 9973076 and by ACS PRF under 
the grant 33851-G5.

\end{document}